\documentclass[]{aa}

\usepackage{txfonts}
\usepackage{graphicx}
\usepackage{natbib}

\begin{document}

\title{A new type of small-scale downflow patches in sunspot penumbrae}

\author{Y. Katsukawa\inst{1} \and J. Jur\v c\'ak\inst{2}}
\institute{National Astronomical Observatory of Japan, 2-21-1 Osawa, 
Mitaka, Tokyo 181-8588, Japan
\and Astronomical Institute of the Academy of Sciences, Fricova
298, 25165 Ond{\u r}ejov, Czech Republic}

\date{Received December 18, 2009; accepted}

\abstract
{Magnetic and flow structures in a sunspot penumbra are created by 
strong interplay between inclined magnetic fields and photospheric
convection. They exhibit complex nature that cannot always be explained 
by the well-known Evershed flow.}
{A sunspot penumbra is observationally examined to reveal properties 
of small-scale flow structures and how they are related to 
the filamentary magnetic structures and the Evershed flow. 
We also study how the photospheric dynamics is related to 
chromospheric activities.} 
{The study is based on data analysis of spectro-polarimetric
observations of photospheric Fe I lines with the Solar Optical Telescope
aboard Hinode in a sunspot penumbra at different heliocentric angles. 
Vector magnetic fields and velocities are derived using the 
spectro-polarimetric data and a Stokes inversion technique. An 
observation with a Ca II H filtergram co-spatial and co-temporal 
with the spectro-polarimetric one is also used to study possible 
chromospheric responses.}
{We find small patches with downflows at photospheric layers. 
The downflow patches have a size of 0.5'' or smaller and have 
a geometrical configuration different from that of the Evershed 
flow. The downflow velocity is about 1 km/s at lower photspheric 
layers, and is almost zero in the upper layers. Some 
of the downflow patches are associated  with brightenings seen in 
Ca II H images.}
{The downflows are possible observational signatures of downward 
flows driven by magnetic reconnection in the interlaced magnetic 
field configuration, where upward flows make brightenings in 
the chromosphere. Another possibility is that they are concentrated
downward flows of overturning magnetoconvection. }

\maketitle

\section{Introduction}
A sunspot penumbra is one of the intriguing structures on the solar
surface, in which filamentary structures are created by photospheric 
convection in the presence of inclined strong magnetic fields. 
Magnetic and flow fields in penumbrae have been studied by many 
authors in order to understand nature of the filamentary structure 
and of the Evershed flow observed at the photospheric layer 
\citep[for reviews, see][]{solanki2003,thomas2004,bellotrubio2010}. 
Observations suggest that the penumbral magnetic fields consist of 
two different components \citep{title1993, solanki1993}: a relatively 
vertical and strong background field; and a weaker and more horizontal 
field where the Evershed flow takes place \citep[e.g.,][]{langhans2005, 
borrero2005, bellotrubio2007}. Recent high resolutional 
spectro-polarimetric observations of penumbrae clearly showed sources 
and sinks of the Evershed flow at the inner and outer ends of penumbral 
filaments \citep{westendorp2001,rimmele2006,ichimoto2007}. Some 
observations indicate existence of elongated upflows at the center 
of the filaments with surrounding downflows \citep{zakharov2008,
rimmele2008} and imply that overturning convection is taking 
place in the filaments. However, observational evidence of downflows 
in penumbrae is still poor at this moment although they are important
to understand the process of heat transport there. 

There is also little knowledge about how the magnetic and flow 
structures in penumbrae dynamically evolve and how they are 
related to chromospheric activities in penumbrae. The solar 
chromosphere is known to have highly dynamic and intermittent 
structures varying on timescales shorter than minutes \citep[see 
recent reviews by][]{rutten2006,wedemeyerbohm2009}. Observations 
with the Hinode satellite have provided new observational evidences 
about fine-scale dynamics of the chromosphere in active regions 
thanks to the stable image quality unaffected by atmospheric distortion 
\citep[e.g.][]{shibata2007}. \citet{katsukawa2007} 
revealed that small-scale transient jet-like brightenings (referred 
to as penumbral microjets) occur ubiquitously in the chromosphere 
above sunspot penumbrae and suggested that magnetic reconnection 
in an interlaced magnetic field configuration is the most plausible
explanation to these penumbral brightenings. It is therefore vital 
importance to accurately measure velocities and vector magnetic 
fields around the penumbral brightenings to ascertain their 
physical origin. 

In this paper, we analyze data sets of spectro-polarimetric 
observations of photospheric spectral lines in a sunspot penumbra 
at different heliocentric angles.  We present newly identified 
small downflow patches with the same magnetic polarity as the spot, 
which are different from the above-mentioned flows associated with 
the Evershed flows. Observations of the penumbral chromosphere 
simultaneous with the spectro-polarimetric data are also analyzed 
to study possible association of chromospheric brightenings with 
the small downflow patches in the photosphere.

\section{Observation and Data Reduction}
\begin{figure*}
\resizebox{\hsize}{!}{\includegraphics{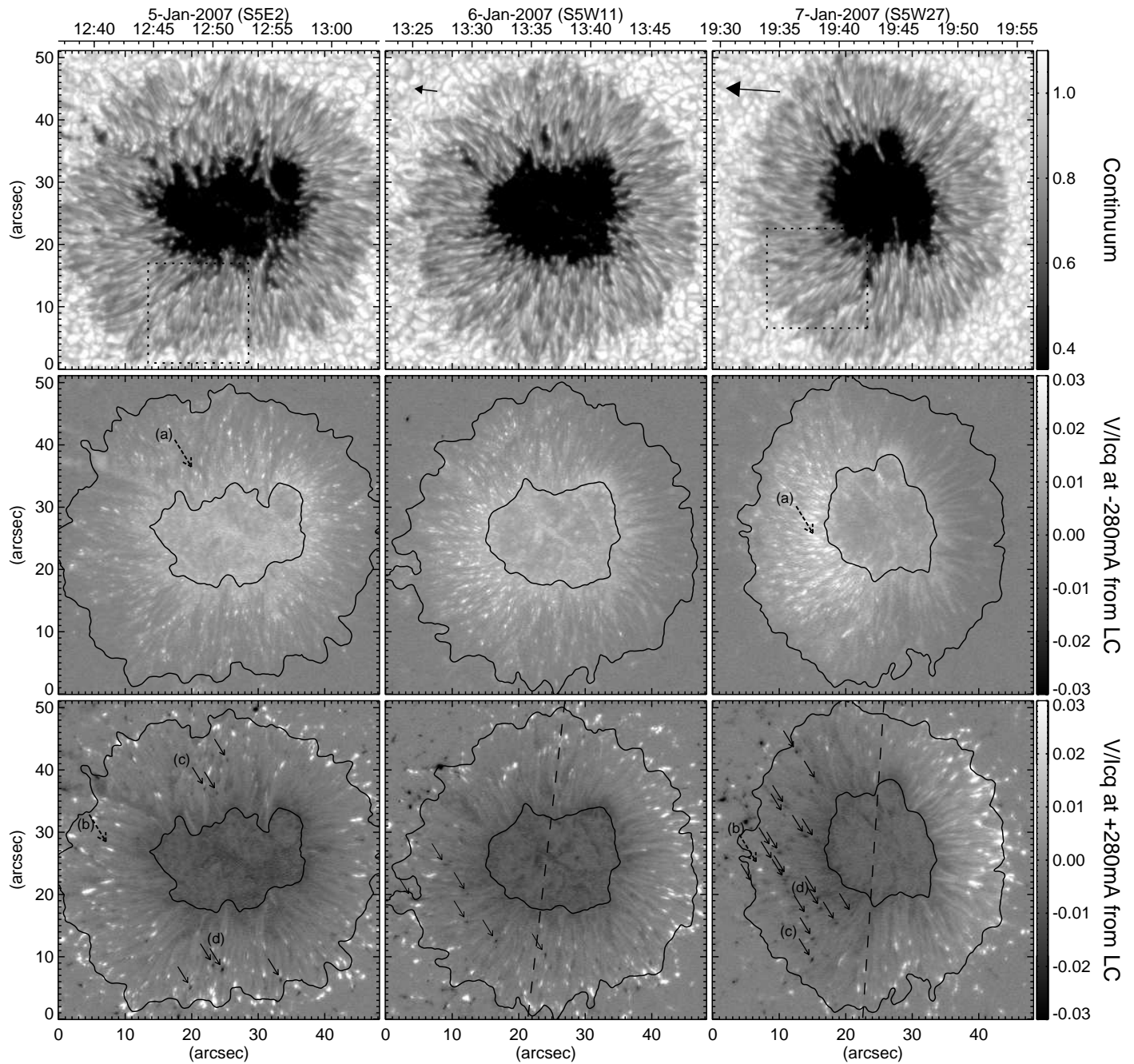}}
\caption{Maps of continuum intensities ({\it top}), Stokes V signals 
$V/I_{cq}$ at the blue wing (-280~m\AA\ from the line center, 
{\it middle}), and Stokes V signals $V/I_{cq}$ at the red wing 
(+280~m\AA\ from the line center, {\it bottom}) of the Fe I 
6301.5~\AA\ line reconstructed from the SP data, where $I_{cq}$ is the 
continuum intensity averaged in the quiet Sun. The dates and hours of 
the SP scanning observations are shown at the top of the maps. The 
arrows on the continuum images ({\it top center} and {\it top right}) 
indicate the direction toward the disk center. The solid arrows on 
the images of Stokes V at the red wing ({\it bottom}) indicate 
locations of small patches having enhanced negative Stokes V signals. 
The arrows labeled with (a) - (d) mark the locations where Stokes 
profiles are shown in Figs. \ref{sprof1} and \ref{sprof2}. The 
two solid contours in the middle and bottom images represent 
boundaries between the umbra and the penumbra and between the 
penumbra and the quiet Sun. The dashed lines on the bottom center 
and bottom right images indicate boundaries separating the disk 
center side and the limb side.}
\label{spmapclv}
\end{figure*}

A sunspot in the active region 10933 was observed on Jan. 5 - 7, 
2007 with the spectropolarimeter (SP) of the Solar Optical Telescope 
\citep[SOT;][]{tsuneta2008, suematsu2008,ichimoto2008b, shimizu2008a}
aboard Hinode \citep{kosugi2007}. We here use data taken with 
three SP scans during the period. The spot was located at heliographic 
coordinates S$5^{\circ}$ E$2^{\circ}$, S $5^{\circ}$ W $11^{\circ}$, and 
S $5^{\circ}$ W $27^{\circ}$ at the time of the SP observations, which are
summarized in Table \ref{tdata}. The SP recorded full Stokes spectra of 
the two Fe I lines at 6301.5 \AA\ and 6302.5\AA\ in a field of view 
(FOV) of about 50''$\times$50'' containing the spot. It took about 
27 minutes with 320 slit steps to map the FOV with a step size of 
0.15'' and an integration time of 4.8~s per slit position. On 7 Jan 
2007, the broadband filter imager (BFI) provided a filtergram 
(FG) observation through the blue continuum (4504 \AA) and Ca II H 
(3968 \AA) filters with a 0.054'' spatial sampling and 30~sec cadence 
simultaneously with the SP observation. Here we concentrate on 
images taken through the Ca II H filter to see chromospheric 
activities. The blue continuum images are used to co-align FG and 
SP data. The small offset between Ca II H and blue continuum images 
is also corrected using the values given in \citet{shimizu2007}. 
Both the SP and BFI data are calibrated with standard routines 
available under Solar Software (SSW). Calibration of wavelength 
positions of the two Fe I lines in the SP data is done using an 
average line-center position outside of the sunspot. The correction 
for the convective blueshift is not applied.

\begin{table}
\caption{\label{tdata}Data sets and the number of the downflow patches in 
the penumbra}
\centering
\begin{tabular}{cccccc}
\hline
\hline
Time & Position & \multicolumn{3}{c}{Number of the patches}\\
\cline{3-5}
  &          & Total & Center & Limb\\
\hline
5-Jan-2007 12:37 - 13:04 & S5$^{\circ}$ E2$^{\circ}$  & 7  & -  & - \\
6-Jan-2007 13:22 - 13:49 & S5$^{\circ}$ W11$^{\circ}$ & 6  & 5  & 1 \\
7-Jan-2007 19:29 - 19:56 & S5$^{\circ}$ W27$^{\circ}$ & 19 & 19 & 0 \\
\hline
\end{tabular}
\end{table}

\begin{figure*}
\centering
\resizebox{\hsize}{!}{\includegraphics{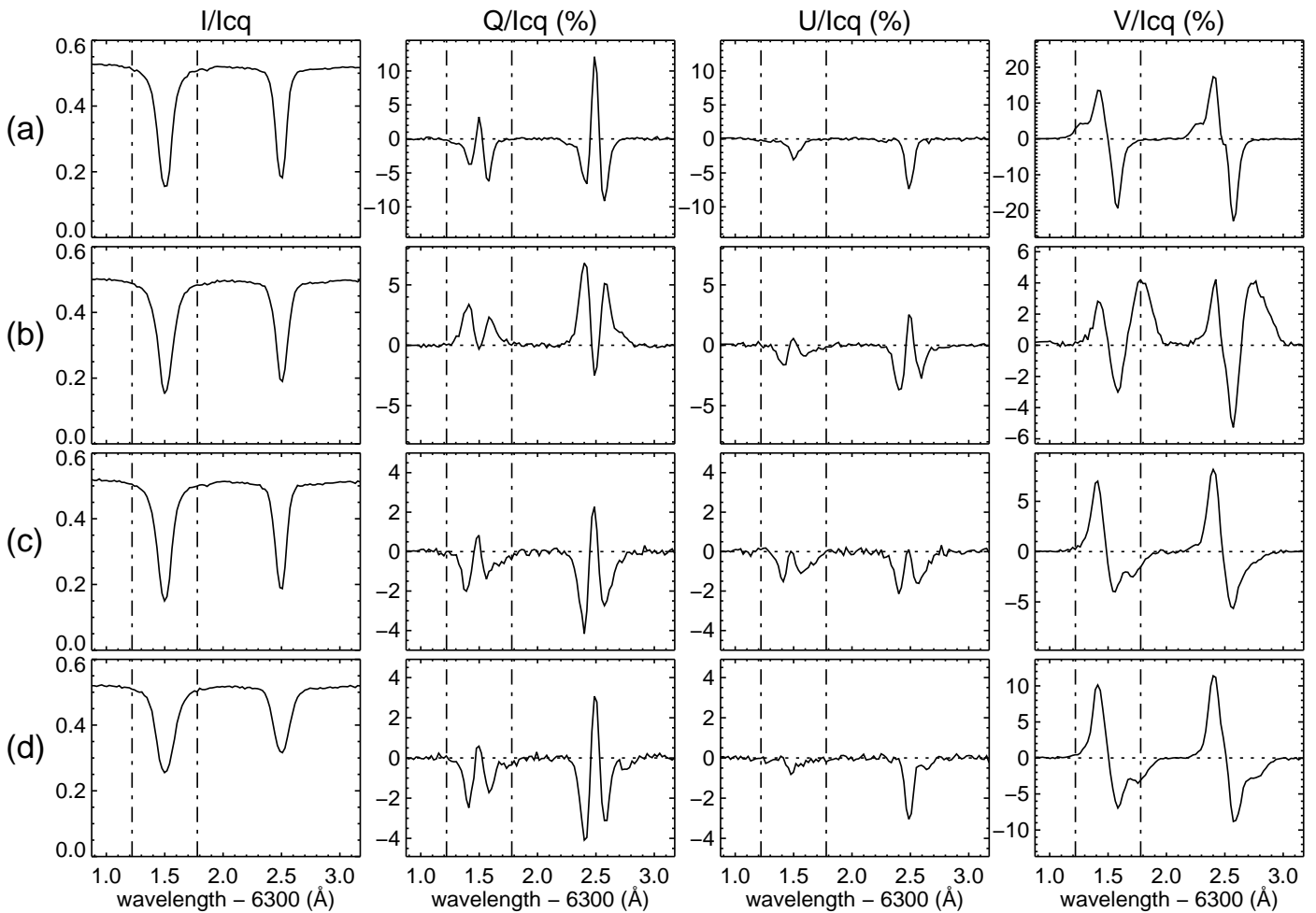}}
\caption{Examples of Stokes profiles seen in the sunspot penumbra 
observed on 5 Jan. 2007 when the spot was located close to the disk 
center. Profiles in the patches with enhanced positive Stokes V 
signals at the blue wing (a), those in the patches with enhanced 
positive Stokes V signals at the red wing (b), and those in the 
patches with enhanced negative Stokes V signals at the red wing 
(c and d). The vertical dash-dotted lines show the wavelength offsets 
of $\pm 280$ m\AA\ from 6301.5 \AA, which are used to make the maps 
of Stokes V signals in Fig. \ref{spmapclv}.}
\label{sprof1}
\end{figure*}

\begin{figure*}
\centering
\resizebox{\hsize}{!}{\includegraphics{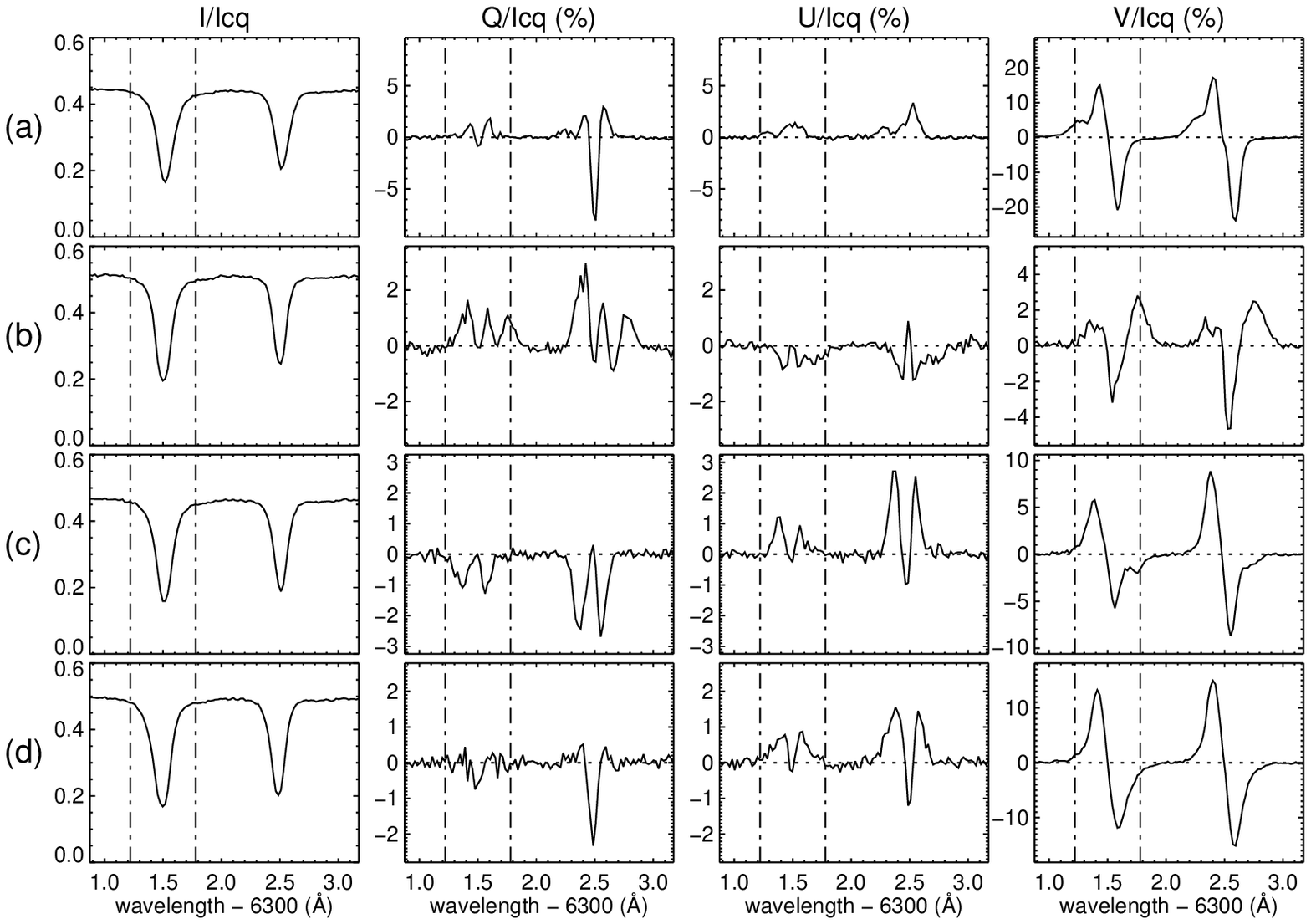}}
\caption{Same as in Fig. \ref{sprof1}, but for 7 Jan. 2007 when the spot
was located away from the disk center.}
\label{sprof2}
\end{figure*}

\section{Results}

\subsection{Patchy Flows in the Penumbra}
Maps of the sunspot constructed from the SP data sets are shown in
Fig.~\ref{spmapclv}. Here are shown maps of continuum intensities
and Stokes V signals at the blue and red wings ($\pm$280~m\AA\ 
from the line center) of the 6301.5~\AA\ line. The Stokes V images at 
far wings have been used to identify Doppler shifts of Stokes V
profiles \citep{ichimoto2007,shimizu2008b,martinez2009}. In the 
sunspot penumbra, especially, many small patches having 
enhancements of Stokes V signals at either wing are commonly 
observed, as is shown in the middle and bottom panels in 
Fig.~\ref{spmapclv}. 

It is realized that there are three kinds of patchy enhancements 
visible in the Stokes V maps of the penumbra. The first one is the 
enhancement at the blue wing. It is easily seen that there are 
many patches with enhanced positive Stokes V signals in the middle 
panels of Fig.~\ref{spmapclv}. They are distributed all over the 
penumbra when the spot is located near the disk center on 5 Jan. while 
they are preferentially found in the center side penumbra when the spot is 
located away from the disk center on 6 and 7 Jan. Examples of Stokes
profiles observed inside the patches indicated by the dashed arrows are 
shown in Figs. \ref{sprof1} (a) and \ref{sprof2} (a). These 
Stokes V profiles are characterized by presence of humps at the blue wing 
with the same polarity as the spot, which indicates presence of 
enhanced blue shifts there. The blue-shifted Stokes V profiles were 
studied by \cite{rimmele2006} and \cite{ichimoto2007}, and it was 
found that they correspond to bright penumbral grains and are the 
inner footpoints of the Evershed flows where hot upflow occurs. 

The second kind of the patchy enhancements in the Stokes V maps is
positive (white) enhancements at the red wing as shown in the bottom 
panels of Fig. \ref{spmapclv}. The positive Stokes V signals at the
red wing are of opposite to the polarity compared to the spot. Examples 
of Stokes profiles with the positive enhancements at the red wing are 
shown in Figs. \ref{sprof1} (b) and \ref{sprof2} (b). The Stokes V 
profiles in these regions are characterized as three lobe profiles which 
are completely different from a regular antisymmetric Stokes V profile. 
The red lobe of the Stokes profiles is produced by a strongly 
red-shifted Stokes V profile with the polarity opposite to the major 
polarity of the spot \cite[e.g.][]{sanchesalmeida2009}. The patches 
with the positive Stokes V signals are mainly observed near the 
boundary of the penumbra when the spot was located near the disk 
center. They are attributed to downflows of the Evershed flows along 
magnetic field lines returning into the photosphere at the penumbral 
boundary \citep{westendorp2001,bellotrubio2004,ichimoto2007,
shimizu2008b}. It is realized that similar positive patches are also 
seen even in the middle of the penumbra in the bottom panels of Fig. 
\ref{spmapclv}. These were studied by \cite{sainzdalda2008} who 
showed that these can be associated with the sea-serpent field 
lines in the mid-penumbra.

The third ones, which have not been realized before, are enhancements 
of Stokes V signals at the red wing similar with the previous one, 
but with the magnetic polarity same as the spot. The patches having
negative enhancements at the red wing are indicated by solid 
arrows in the bottom panels of Fig. \ref{spmapclv}. The patches are
less frequent than the patches with the positive enhancements, and 
their sizes are smaller than 0.5 arcseconds. This is probably 
why the features have not been realized before. Figs. \ref{sprof1}, 
\ref{sprof2} (c) and (d) show examples of Stokes profiles 
observed inside the patches marked with the solid arrows. The 
Stokes V profiles have humps at the red wing with the same 
polarity. The Stokes V profile in Figs. \ref{sprof2} (d) does 
not have a hump at the red wing, but it has a red tail causing 
the enhancements in the red wing images of Stokes V, and the Stokes 
V amplitude of the red lobe is smaller than that of the blue lobe. 
The enhancements at the red wing are not so large in the Stokes Q 
and U profiles though there are weak tails at the red wing in the
Stokes Q profiles in Fig. \ref{sprof1} (c) and (d). The Stokes 
profiles are easily distinguishable from the known profiles which 
are supposed to be associated with the Evershed flow. The enhanced 
Stokes V signals at the red wing can be caused by either 
temperature changes or magnetic field differences or Doppler 
shifts. But the asymmetric profiles suggest that the main cause 
of the enhancements are red shifts of Stokes V profiles because 
the former two factors are expected to influence both the red and 
blue wings evenly.

\subsection{Spatial Distribution of the Downflow Patches}

We here refer the patchy regions with enhanced Stokes V signals of 
the same polarity at the red wing as the downflow patches hereafter.
We identify the downflow patches in the penumbra with a criteria 
where $V/I_{cq}$ are smaller than -0.015 at -280~m\AA\ from 6301.5\AA.
The downflow patches thus identified are indicated by the solid arrows 
in the bottom panels of Fig. \ref{spmapclv}. The downflow patches tend
to be seen in the middle to outer penumbra. In addition, it is clear
that most of the patches are observed in the disk center side penumbra 
when the sunspot is located away from the disk center while the positive 
patches at the red wing are more frequently seen in the limb side 
penumbra. Table~\ref{tdata} shows the number of the downflow patches 
identified in the penumbra observed in the three SP scans taken when 
the sunspot was located at different angles from the disk center. It 
is clear that the number of the patches identified is largest on 7 Jan 
when the spot was located about 30$^{\circ}$ away from the disk center 
while the size of the spot is not so different. Also most of the 
downflow patches are seen in the disk center side penumbra when 
the spot is not located near the disk center on 6 and 7 Jan.

\begin{figure*}
\resizebox{\hsize}{!}{\includegraphics{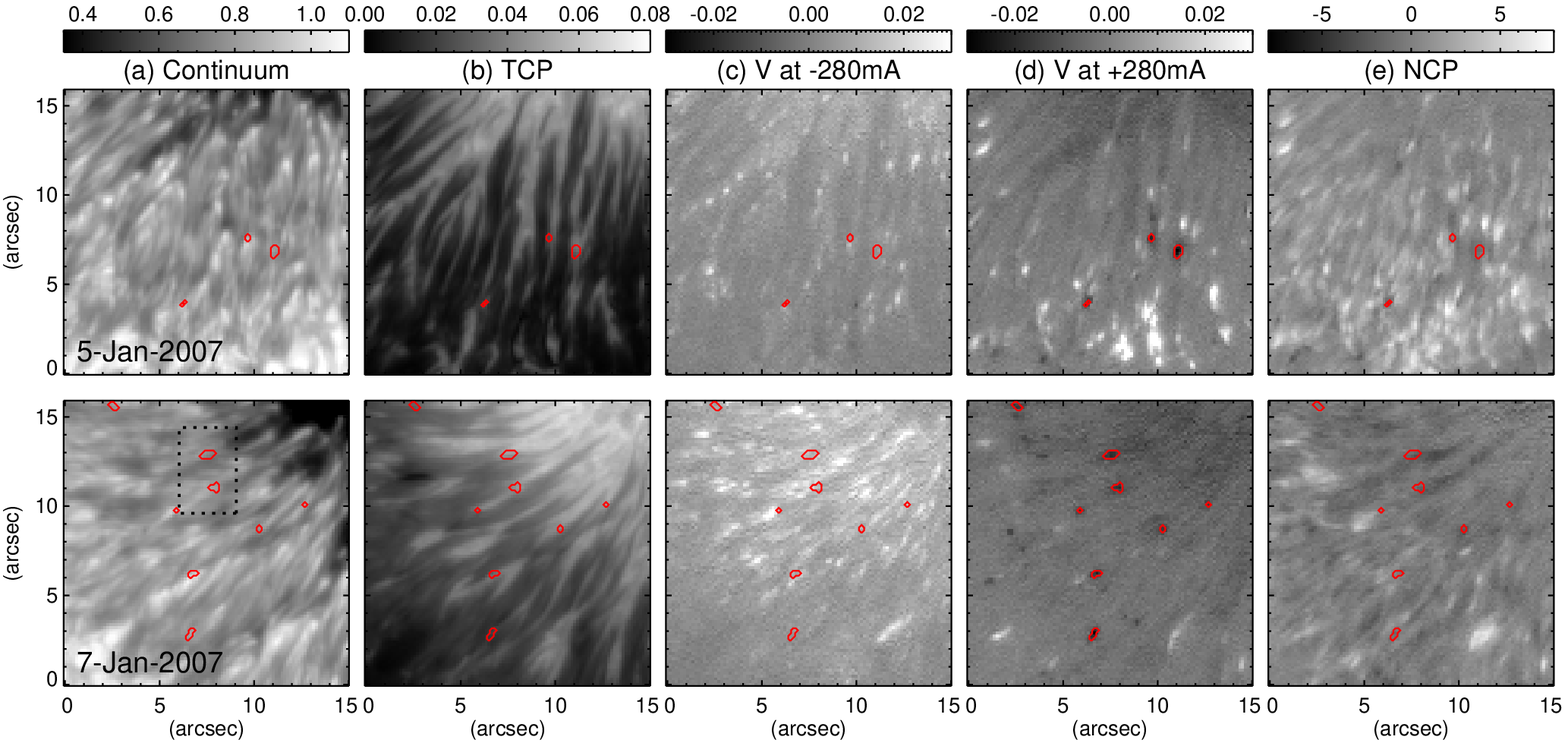}}
\caption{(a) SP continuum images observed on 5 and 7 Jan. in the 
region indicated by the dashed boxes in Fig. \ref{spmapclv}, (b) 
total circular polarization (TCP, $\int |V| d\lambda/\int I_{c} d\lambda$) 
integrated over $\pm500$ m\AA\ around the line center of 6301.5~\AA\ line, 
(c) Stokes V signals $V/I_{cq}$ at -280 m\AA\ from the line center of 
the 6301.5~\AA\ line, (d) Stokes V signals $V/I_{cq}$ at +280~m\AA\ 
from the line center of the 6301.5~\AA\, and (e) net circular 
polarization (NCP, $\int V d\lambda/I_{c}$, in the unit of m\AA) integrated 
over $\pm500$ m\AA\ around the line center of the 6301.5~\AA\ line,
where $I_{cq}$ is the continuum intensity averaged in the quiet Sun
and $I_{c}$ is the local continuum intensity. The contours on each 
panel represent $V/I_{cq}=-0.015$ at +280~m\AA\ from 6301.5\AA. The 
dotted boxes in (c) -- (f) indicate the region of interest used 
in Fig.~\ref{velmap}. }
\label{spmap}
\end{figure*}

Due to the presence of the downflow patches, we study in detail the 
regions observed on 5 and 7 Jan., which are indicated by the dotted 
boxes in the top rows of Fig.~\ref{spmapclv}. The SP images of the 
regions are shown in the panels (a) -- (e) of Fig.~\ref{spmap}. The 
contours in Figs.~\ref{spmap} indicate $V/I_{cq}=-0.015$ at +280~m\AA\ 
to locate the downflow patches. The size of the patches is 0.5'' or 
smaller and the patches are mostly found in bright penumbral 
filaments and in regions with enhanced total circular polarization 
(TCP) as is shown in Fig.~\ref{spmap} (a) and (b). Some of the downflow 
patches are found to be associated with penumbral grains, which are 
bright structures at the inner end of penumbral filaments, as can be 
seen in Fig.~\ref{spmap} (a). Such patches are located near the regions
with enhanced positive V signals at the blue wing (Fig.~\ref{spmap} [c]), 
which is expected since the penumbral grains are found to be 
associated with the Evershed upflows \citep{rimmele2006,ichimoto2007}. 
But the locations of the downflow patches are slightly off from the 
upflow patches as can be seen in Fig.~\ref{spmap} (c). The downflow 
patches tend to have small negative net circular polarization (NCP) 
in Fig.~\ref{spmap} (e). This is expected because the Stokes V profiles
have negative enhancements at the red wing in the downflow patches 
(see Fig.~\ref{sprof1} [c], [d] and \ref{sprof2} [c], [d]). 
The NCP, which is the Stokes V signal integrated over the spectral 
line, is used to quantify the asymmetries in Stokes V profiles which 
are generated by gradients in velocities and magnetic fields along 
the line-of-sight (LOS) in the line forming region \citep{illing1975, 
auer1978}. \cite{ichimoto2008a} found that there are structures with 
non-zero NCP observed in inter-Evershed flow lanes. The downflow patches 
partially contribute to make the non-zero NCP in the lanes because 
the downflow patches are mostly observed there. But we cannot see 
clear spatial correlation between the structures seen in the NCP 
images and the downflow patches. 

\subsection{Velocity and Magnetic Fields}
\begin{figure*}
\centering
\includegraphics[width=18cm]{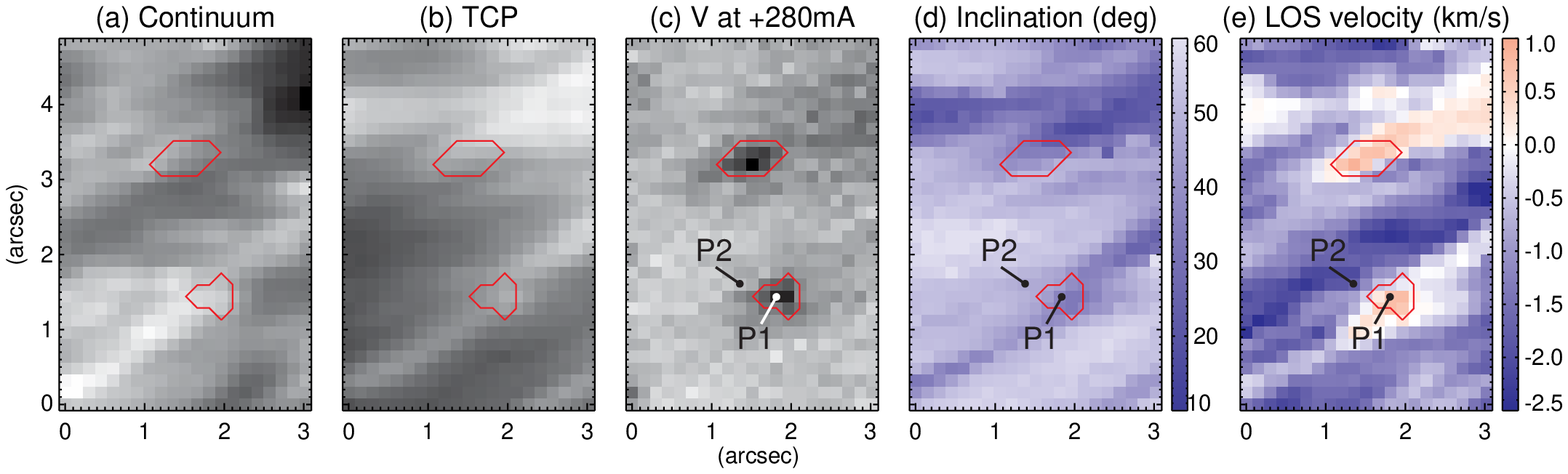}
\caption{Closeup view of the downflow patches in the penumbra. 
Here are shown (a) a map of continuum intensities, (b) total 
circular polarization (TCP), (c) Stokes V signals at +280~m\AA\ 
from the line center of the 6301.5~\AA\ line, (d) magnetic 
field inclination angles from the local normal, and (e) LOS 
velocities derived by the SIR inversion, where positive 
(negative) velocities represent motion away from (toward) 
the observer. The inclination angles and the velocities are
averaged ones in optical depths from $\log(\tau_c)=-0.3$ to 
$\log(\tau_c)=-0.5$. P1 and P2 in (c) -- (e) indicate pixels 
where the Stokes V profiles shown in Fig. \ref{vprof} (a) were 
observed. The contours on each panel represent $V/I_{cq}=-0.015$ 
at +280~m\AA\ from 6301.5\AA.}
\label{velmap}
\end{figure*}

\begin{figure}
\resizebox{\hsize}{!}{\includegraphics{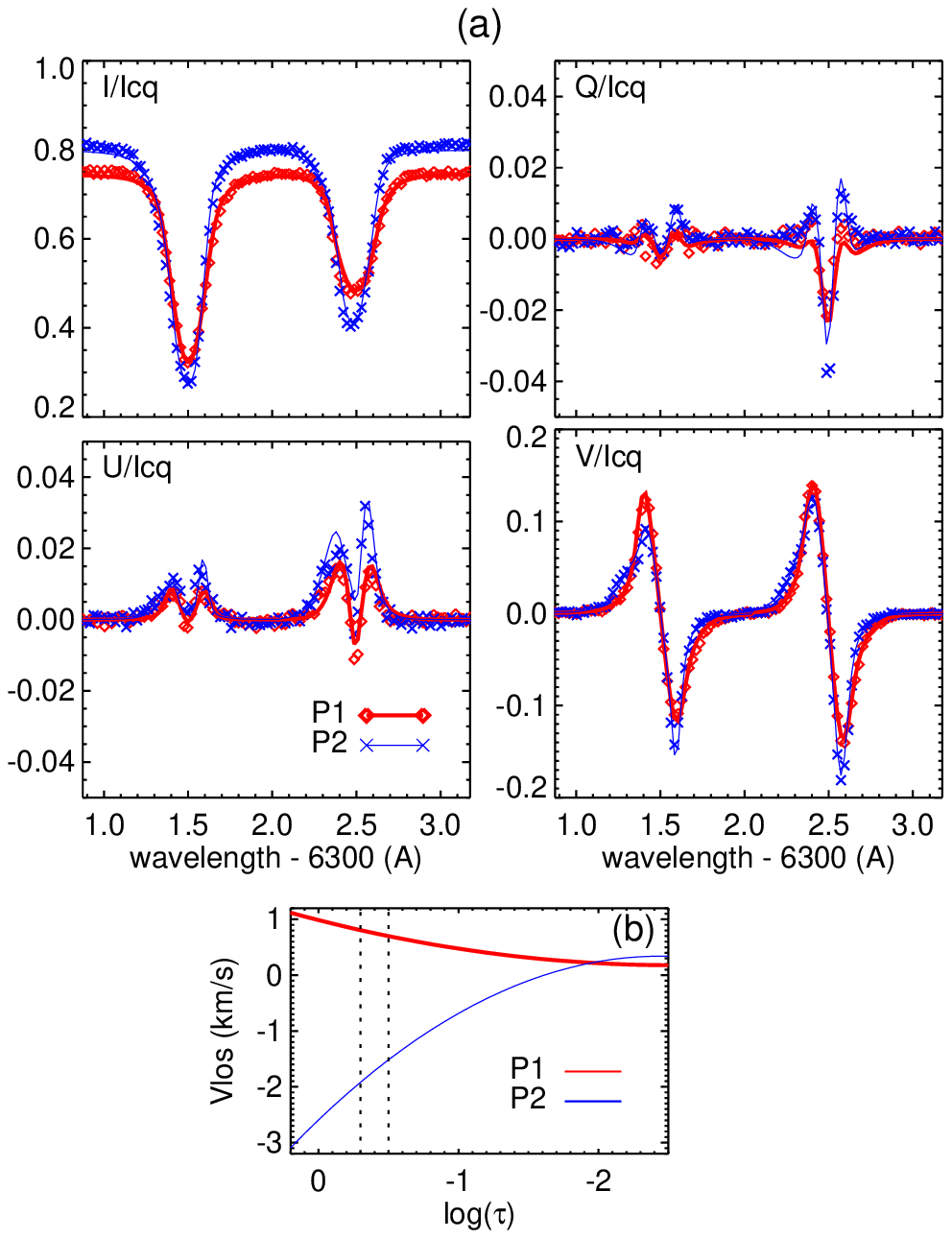}} 
\caption{(a) Stokes profiles observed inside (P1) and outside (P2) 
of the downflow patch. The diamonds and crosses show observed Stokes 
profiles at P1 and P2, respectively. The red and blue solid curves 
indicate best-fit profiles from the SIR inversion at P1 and 
P2, respectively. (b) The stratifications of LOS~velocity derived by 
the SIR inversion at P1 and P2. Positive (negative) velocities represent 
motion away from (toward) the observer.}
\label{vprof}
\end{figure}

To get vector magnetic fields and line-of-sight (LOS) velocities, we 
use the inversion code SIR \citep[Stokes Inversion based on Response 
function;][]{ruizcobo1992} in the small region indicated by the dotted 
box on the continuum image taken on 7 Jan in Fig. \ref{spmap}. 
Figure~\ref{velmap} (a) -- (c) shows a closeup of the region in 
continuum intensities, total circular polarization (TCP), and Stokes 
V signals at +280 m\AA. This enlarged area contains two typical 
downflow patches that we found in our data. Because the 
observed Stokes profiles in the downflow patches have asymmetries 
as shown in Fig.~\ref{sprof1} and ~\ref{sprof2}, height dependence of 
plasma parameters is taken into account in the inversion. We also 
consider an effect of polarized (i.e. not only Stokes I but also 
Q, U, and V) stray lights by using an averaged Stokes profile from 
surrounding pixels ($13\times13$ pixels area) for each pixel 
\footnote{For a study of the stray light contamination in Hinode 
SP data, see \cite{orozco2007}.}, because it provides better fitting 
at most of the pixels. The stray-light fraction is fixed to 15\% in the 
inversion. We found that changes of the stray-light fraction have 
little influence on resulting magnetic fields and Doppler velocities 
in this analysis. We use three nodes in optical depth for temperature,
velocity, and magnetic field inclination and azimuth and five nodes for
magnetic field strength. The final stratifications are obtained by 
parabolic (for three nodes) and spline (for five nodes) interpolation
across those nodes. In order to obtain field inclination 
angles from the local normal, the 180 degrees ambiguity in the LOS 
azimuth angle has to be resolved because the sunspot was located away 
from the disk center. It can be done without any difficulties in this 
analysis, because we can assume that magnetic field orientation is 
radially and smoothly directed in the penumbra. Fig.~\ref{vprof} (a) 
shows Stokes profiles observed in the downflow patch (P1, which is 
identical with the profile shown in Fig.~\ref{sprof2} [d]) and a 
typical profile observed in the center-side penumbra with the signature
of the Evershed flow (P2), and best-fit profiles from the SIR inversion 
of the observed profiles. The locations of P1 and P2 are indicated in 
the panels (c) -- (e) of Fig.~\ref{velmap}.

As is expected from the observed Stokes V profiles 
(Fig.~\ref{sprof1} [c], [d], Fig.~\ref{sprof2} [c], [d], 
and Fig.~\ref{vprof} [a]), the biggest difference in LOS velocities is 
expected at the lower photospheric layers since the profiles P1 and P2 
differ significantly at the far wings, but similar velocities ought to 
be obtained at higher layers as the Stokes V zero-crossing wavelengths 
of the shown profiles are the same. This is confirmed by the inversion 
as we obtain the highest difference between the velocity stratifications 
at lower layers (Fig.~\ref{vprof} [b]). There is a flow of about 
1 km/s away from the observer at low photospheric layers of the 
downflow patch (P1) while a flow up to $\sim$ 3 km/s toward the 
observer is observed in P2, which is attributed to the Evershed 
flow in the center-side penumbra. Panels (d) and (e) in 
Fig.~\ref{velmap} show the spatial distribution of magnetic inclination 
angles from local normal and LOS velocities, where the inclinations 
and the velocities represent average values of the parameters derived 
by the SIR inversion in the range of optical depths from 
$\log(\tau_c)=-0.3$ to $\log(\tau_c)=-0.5$. The LOS velocities are 
around zero in bright penumbral filaments where the inclination angles 
are relatively vertical. The small areas with enhanced negative Stokes 
V signals at the red wing have positive velocities, which 
indicates presence of a flow away from the observer. Elsewhere, 
we found negative LOS velocities caused by the Evershed flow. 

\subsection{Chromospheric Brightenings}

\begin{figure*}[t]
\centering
\includegraphics[width=17cm]{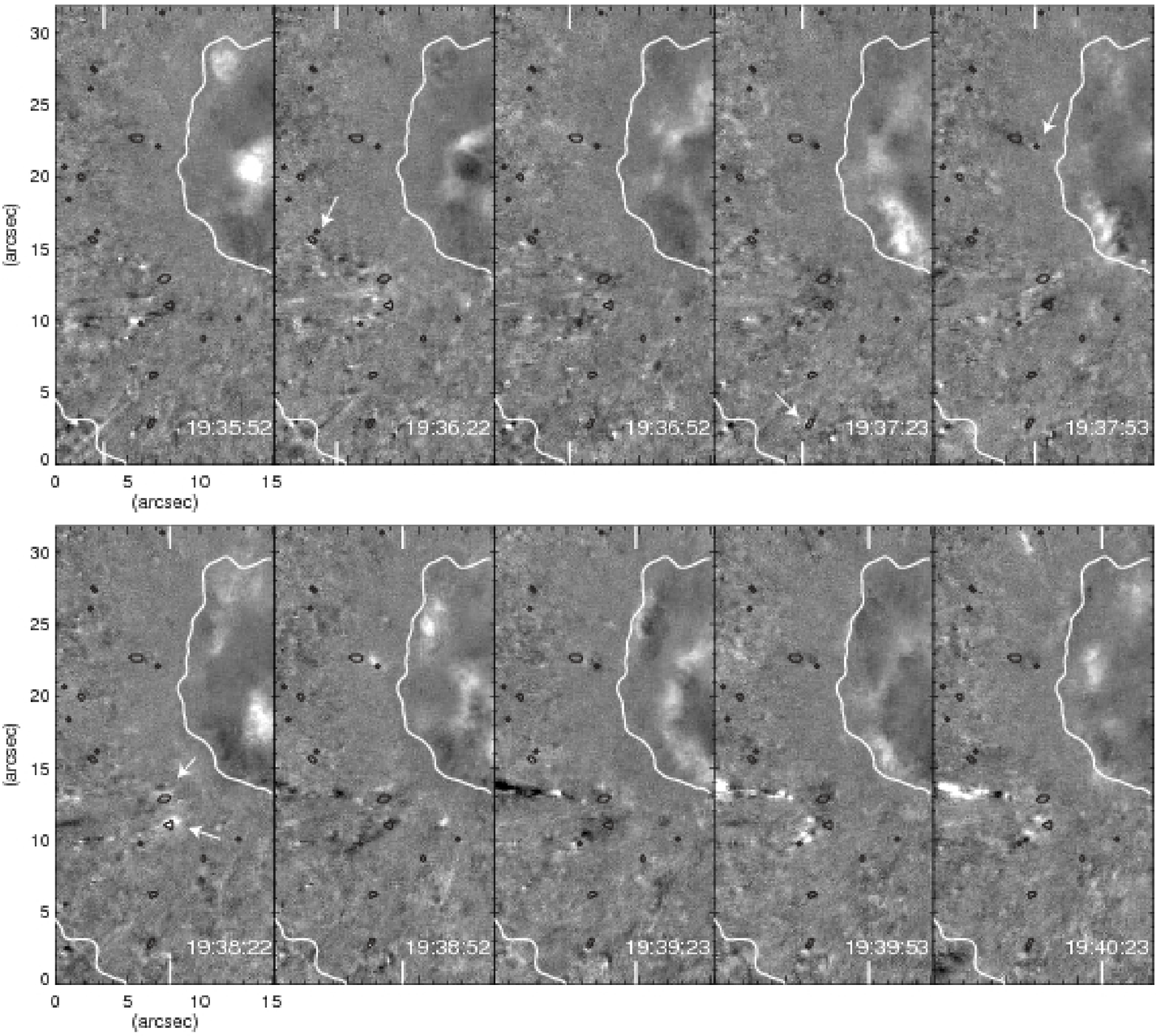} 
\caption{FG Ca II H images processed with the temporal high-pass 
Fourier filtering. The images were obtained between 19:35:52 and 
19:40:23 on 7 Jan. 2007 simultaneous with the SP scan. The FOV 
covers the center-side penumbra. The solid contours show 
$V/I_{cq}=-0.015$ at +280~m\AA\ from 6301.5~\AA\ indicating the 
downflow patches as in Fig.~\ref{spmap}. The white contours on 
each panel represent the boundaries between the umbra and the 
penumbra and between the penumbra and the quiet Sun. 
The thick white vertical lines on each Ca II H image indicate 
positions of the SP slit at each Ca II H exposure.}
\label{camap}
\end{figure*}

In order to study a possible counterpart in the chromosphere 
associated with the photospheric downflow patches in the penumbra, 
we use the data set taken on 7 Jan 2007 in which Ca II images were
taken simultaneous with the SP scan. A temporal high-pass Fourier 
filtering is applied at each spatial pixel with a cut-off frequency 
of 5~mHz to extract transient chromospheric brightenings seen in 
the Ca II H images. The images thus processed are shown in 
Fig.~\ref{camap}. Only the center side penumbra is shown because 
the downflow patches are preferentially observed there. Transient 
brightenings are easily identified in the image sequence. Most of 
them are point-like, but jet-like elongated brightenings can also be 
identified. Their lifetimes are typically comparable to or 
shorter than the temporal cadence (30~sec) of the FG observation. 
The penumbral microjets were found to roughly follow the orientation 
of background magnetic fields in the penumbra, and the inclination 
angle is from 40$^{\circ}$ to 60$^{\circ}$ with respect to the 
local vertical in the mid-penumbra \citep{katsukawa2007, jurcak2008}. 
The spot studied here was located at about 30$^{\circ}$ away from the 
disk center in the observation. This means that the orientation of 
the microjets tend to be aligned with the line-of-sight direction 
in the center-side penumbra, and these might be observed as point-like 
brightenings there.

When comparing the FG and SP observations, we have to be careful about 
co-alignment between the SP slit position and the FG field of view, 
especially when we are interested in short-lived phenomena. For 
this reason, the slit position at each Ca II H exposure is indicated 
in Fig.~\ref{camap} by thick vertical lines. We focus on Ca II H
brightenings temporally and spatially coincident with the SP slit 
positions. Two brightenings are clearly observed in the Ca II H 
image taken at 19:38:22 around the SP slit position. These brightenings
correspond to two downflow patches (represented by contours in 
Fig.~\ref{camap}) that are indicated by the arrows. There are 
another faint brightenings observed near downflow patches and the 
SP slit position at 19:36:22, 19:37:23, and 19:37:53, again indicated 
by the arrows. Thus it appears that at least some downflow patches are 
spatially and temporally related to chromospheric brightenings. 
A higher-cadence Ca II H image sequence are needed to investigate 
this possibility in more detail.

\section{Summary and Discussion}
Using the Hinode spectro-polarimetric observations, we found 
photospheric downflows that have the same magnetic polarity as the 
sunspot umbra. The Stokes profiles seen in the patches are 
characterized as humps or enhancements at the red wing of the Stokes
V profiles. Probably due to the small size (smaller than 0.5''), 
such downflow patches have not been reported previously in penumbrae. 
They are preferentially found inside the bright penumbral filaments 
and in the center-side penumbra in our data. It was also found 
that some of the downflow patches are temporally and spatially 
coincident with chromospheric brightenings in the Ca II H images.

As is shown in Fig.~\ref{spmapclv} and Fig.~\ref{spmap}, the number 
of the downflow patches observed in the penumbra is not so many compared 
with the structures associated with the Evershed flows. This might 
be related to temporal evolution of the downflow patches, although 
we cannot determine their lifetimes from the single SP scan. In 
order to observe a short-lasting downflow in SP data, the slit must 
cross the downflow patch just at the time when it occurs. This would 
make the detection in a SP observation less frequent. Temporal evolution 
of the downflow patches in penumbrae is studied in our another paper 
\citep{jurcak2010}. The difference of the spatial distribution 
between the limb-side and the center-side when the spot was located away 
from the disk center can be simply explained by orientation of the 
downflows and a projection effect. If the downflows follow field 
lines of the stronger and more vertical magnetic component \citep[as 
suggested by][]{jurcak2008}, the inclination angle of the downflow is 
around 40$^{\circ}$ from local normal in the mid-penumbra 
(Fig.~\ref{velmap} [d]). In this case, the downflows tend to be parallel 
to the LOS direction in the center-side penumbra while they are almost 
perpendicular to the LOS direction in the limb-side penumbra. This means 
that the downflow patches have small LOS velocities in the limb-side 
penumbra, and therefore they cannot be distinguished from the other 
flow structures. When the spot is located away from the disk center, 
not only a vertical motion but a horizontal one make partial 
contribution to the LOS velocities observed as Doppler shifts.
Penumbral grains are known to move toward an umbra with velocities of 
about 0.5~km/s \citep[e.g.][]{muller1973}. When the motion of penumbral 
grains is not apparent but is associated with true mass motion, their 
LOS components are smaller than 0.3~km/s when the heliocentric angle 
is 30 degrees, and cannot completely explain the red shifts seen in 
the downflow patches. In the case when the spot is close the 
disk center, the horizontal motion of penumbral grains does not 
significantly contribute to the Doppler shifts while we can observe 
the downflow patches there. This also supports that the downflows are 
not the result of the motion of penumbral grains.

A possible mechanism of the downflows is that they are driven by a
downward outflow from magnetic reconnection, while an upward one 
cause chromospheric brightenings seen in the Ca II H images. Recently 
\cite{sakai2008} and \cite{magara2010} studied magnetic reconnection 
between horizontal and vertical magnetic fields in a penumbra using 
a numerical simulation, and demonstrated bidirectional 
flows propagating along vertical magnetic fields as a result of 
redirection of outflows along the vertical magnetic field. 
\cite{ryutova2008} proposed that some of the chromospheric brightenings 
are produced by shocks resulted from a sling-shot effect associated 
with magnetic reconnection process in neighboring penumbral filaments. 
The velocity stratification shown in Fig.~\ref{vprof} (b) have a 
possibility to provide useful information to know where magnetic 
reconnection takes place if the downflow is driven by this mechanism. 
The downflow in the lower photosphere suggests that a reconnection 
site might be located in the photosphere because significant 
downflows ought to be observed even in the upper and middle 
photosphere if magnetic reconnection occurs in the chromosphere. 
Because we used only three nodes along the optical depth in the
Stokes inversion and derived the velocity stratification among the 
nodes with the interpolation, we cannot argue acceleration or 
deceleration of the downward flow from the inversion. Further 
spectroscopic studies are necessary to resolve the height dependence 
of their flow structures from the middle to the upper photosphere.

Another possibility of the patchy downward flows are indications of 
convective rolls in penumbral filaments as proposed by \cite{zakharov2008} 
who found blueshifts on the limbward side of a penumbral filament, and 
weak red shifts on the centerward side. It is, however, not clear how 
the convective rolls make sporadic downflow patches. \cite{ortiz2010} 
found small concentrations of downflows associated with strong upflows 
in umbral dots. Their sizes smaller than 0.5'' and lifetimes shorter 
than a few minutes look similar with the nature of the downflow patches 
in the penumbra studied in this paper. There is a possibility that 
magnetoconvection in the presence of strong magnetic fields 
transiently makes small concentration of downflows in the lower
photosphere, which can be explored with numerical simulations 
recently developed \citep[e.g.][]{rempel2009}.

\begin{acknowledgements}
The authors would like to thank the anonymous referee for its helpful 
comments and suggestions. Hinode is a Japanese mission developed and 
launched by ISAS/JAXA, with NAOJ as domestic partner and NASA and STFC 
(UK) as international partners. It is operated by these agencies in 
co-operation with ESA and NSC (Norway). We thank D. Orozco Su\'arez 
for helpful comments. J. J. was supported by a Research Fellowship 
from the Japan Society of the Promotion of Science for Young 
Scientists. Financial support from GA AS CR IAA300030808 is 
gratefully acknowledged.
\end{acknowledgements}

\end{document}